\documentclass[preprint,aps,showpacs,nofootinbib,preprintnumbers,amsmath,amssymb,superscriptaddress]{revtex4-1}
\usepackage{amssymb}
\usepackage{epsfig}
\usepackage{latexsym}
\usepackage{graphicx}
\usepackage{subfigure}
\usepackage{dcolumn}
\usepackage{bm}
\usepackage[usenames ,dvipsnames]{xcolor}
\usepackage[utf8]{inputenc}
\usepackage{slashed}
\usepackage{cleveref}

\begin{document}
\title{Contributions to the Muon $g-2$ from a Three-Form Field}
\author{Da~Huang\footnote{dahuang@bao.ac.cn}}
\affiliation{National Astronomical Observatories, Chinese Academy of Sciences, Beijing, 100012, China}
\affiliation{School of Fundamental Physics and Mathematical Sciences, Hangzhou Institute for Advanced Study, UCAS, Hangzhou 310024, China}
\affiliation{International Centre for Theoretical Physics Asia-Pacific, Beijing/Hangzhou, China}
\author{Yong~Tang\footnote{tangy@ucas.ac.cn}}
\affiliation{University of Chinese Academy of Sciences (UCAS), Beijing 100049, China}
\affiliation{School of Fundamental Physics and Mathematical Sciences, Hangzhou Institute for Advanced Study, UCAS, Hangzhou 310024, China}
\affiliation{National Astronomical Observatories, Chinese Academy of Sciences, Beijing, 100012, China}
\affiliation{International Centre for Theoretical Physics Asia-Pacific, Beijing/Hangzhou, China}
\author{Yue-Liang~Wu\footnote{ylwu@ucas.ac.cn}}
\affiliation{Institute of Theoretical Physics, Chinese Academy of Sciences, Beijing 100190, China}
\affiliation{School of Fundamental Physics and Mathematical Sciences, Hangzhou Institute for Advanced Study, UCAS, Hangzhou 310024, China}
\affiliation{International Centre for Theoretical Physics Asia-Pacific, Beijing/Hangzhou, China}
\affiliation{University of Chinese Academy of Sciences (UCAS), Beijing 100049, China}

\date{\today}
\begin{abstract}
\noindent We examine contributions to the muon dipole moment $g-2$ from a 3-form field $\Omega$, which naturally arises from many fundamental theories, such as the string theory and the hyperunified field theory. In particular, by calculating the one-loop Feynman diagram, we have obtained the leading-order $\Omega$-induced contribution to the muon $g-2$, which is found to be finite. Then we investigate the theoretical constraints from perturbativity and unitarity. Especially, the unitarity bounds are yielded by computing the tree-level $\mu^+\mu^-$ scattering amplitudes of various initial and final helicity configurations. As a result, despite the strong unitarity bounds imposed on this model of $\Omega$,  we have still found a substantial parameter space which can accommodates the muon $g-2$ data.
\end{abstract}

\maketitle

\section{Introduction}\label{s1}
A three-form field, {\it i.e.}, an antisymmetric rank-three tensor field, can generically appear in many fundamental theories trying to unify the General Relativity and Quantum Field Theory. For example, in the $10$-dimensional Type-IIA superstring and supergravity theory, a 3-form field occurs as an important constituent in the Ramond-Ramond sector~\cite{Green:1987sp,Green:1987mn,Polchinski:1998rr}. Furthermore, in the recently proposed Hyper-Unified Field Theory (HUFT)~\cite{Wu:2015wwa,Wu:2017rmd, Wu:2017urh, Wu:2018xah, Wu:2021ign,Wu:2021ucc,Wu:2022mzr}, such a 3-form field is naturally identified as a part of the spin gauge field in this 19-dimensional theory. {Unfortunately, the 3-form field in the superstring theory couples to the ordinary matter with the gravitational strength~\cite{Green:1987sp,Green:1987mn,Polchinski:1998rr}, so it is difficult to be visible at low energies. On the other hand, in the HUFT, by dynamically breaking the conformal symmetry~\cite{Wu:2017urh,Wu:2021ign,Wu:2021ucc} and compactifying the theory into our mundane 4-dimensional spacetime, we expect that a light 3-form field $\Omega$ is left and potentially generates new observable effects~\cite{Sezgin:1980tp}. }  

In this study, we consider the indirect loop effects induced by a 3-form field on the muon electromagnetic dipole moment defined by $a_\mu = (g-2)_\mu/2$, which might solve the long-standing muon $g-2$ anomaly~\cite{Workman:2022ynf}. Currently, the comparison between the Standard Model (SM) prediction and the experimental data shows a remarkable $\sim 4.25\sigma$ discrepancy~\cite{Muong-2:2021ojo}:
\begin{eqnarray}
	\Delta a_\mu = a_\mu^{\rm Exp} - a_\mu^{\rm SM} = (251\pm 59)\times 10^{-11}, 
\end{eqnarray}
where the experimental value $a_\mu^{\rm Exp} = (116\,592\,061\pm 41)\times 10^{-11} $ is obtained by combining the earlier Brookhaven~\cite{Muong-2:2006rrc} and the latest Fermilab Muon $g-2$~\cite{Muong-2:2021ojo} data, while the SM prediction is given by $a^{\rm SM}_\mu = (116\,591\,810\pm 43)\times 10^{-11}$ based on the state-of-art evaluations of various contributions~\cite{Aoyama:2012wk,Aoyama:2019ryr,Czarnecki:2002nt,Gnendiger:2013pva,Davier:2017zfy,Keshavarzi:2018mgv,Colangelo:2018mtw,Hoferichter:2019mqg,Davier:2019can,Keshavarzi:2019abf,Kurz:2014wya,Melnikov:2003xd,Masjuan:2017tvw,Colangelo:2017fiz,Hoferichter:2018kwz,Gerardin:2019vio,Bijnens:2019ghy,Colangelo:2019uex,Blum:2019ugy,Colangelo:2014qya} (see {\it e.g.} Ref.~\cite{Aoyama:2020ynm} for a recent review and references therein). In light of these precise measurements and calculations, this muon $g-2$ anomaly is usually regarded as an indication to the new physics beyond the SM. In the literature, there have already been many new physics attempts to resolve this anomaly (see {\it e.g.}, Ref.~\cite{Athron:2021iuf} for a recent review and references therein). It is worth to mention that, in the HUFT, the 3-form field $\Omega$, as a part of the spin gauge field, couples to the SM fermionic fields inevitably. Hence, it is highly probable to generate an extra contribution to the muon $g-2$ via its interaction with muons. Here we would like to carefully examine if the one-loop contributions induced by $\Omega$ could help us to ameliorate or even eliminate the $(g-2)_\mu$ discrepancy. Furthermore, the scenario we are considering needs a relatively light $\Omega$ and a moderately large coupling to muons, which might potentially lead to the breakdown of perturbativity and unitarity. For that reason, we shall also take into account theoretical constraints from these issues in the framework of effective field theory. 

The paper is organized as follows. In Sec.~\ref{SecLag}, we present the effective field theory framework of the 3-form field $\Omega$. In particular, we show the Lagrangian and the propagator of $\Omega$, together with a short discussion of the significance of the $\Omega$ mass term. Then the leading-order contribution to the muon $g-2$ induced by $\Omega$ is given in Sec.~\ref{SecG2} by calculating the associated one-loop Feynman diagram. Sec.~\ref{SecUnitary} is devoted to studying the theoretical constraints from perturbativity and unitarity on our 3-form model. The numerical results are shown in Sec.~\ref{SecResult} by carefully examining the parameter space of interest. Finally, we conclude in Sec.~\ref{SecConclusion}, together with a short discussion of the possible collider signatures of the 3-form field.


\section{Lagrangian and Conventions}\label{SecLag}
In this section, we set up our notations and conventions for our model of the 3-form field $\Omega = \Omega_{\mu\nu\rho}\,dx^\mu \wedge dx^\nu \wedge dx^\rho$. In the fundamental theories such as the Type-IIA superstring theory or the HUFT, $\Omega$, as a gauge field, couples universally to other fields or particles. However, these fundamental theories are generically living in the spacetime with dimensions higher than 4. In order to explain the physics of our ordinary 4-dimensional spacetime world, one usually involves the compactification technique to wrap the extra dimensions in small compact spaces~\cite{Polchinski:1998rr,Green:1987sp,Green:1987mn}. In this process, due to the wavefunctions of various fields in the extra dimensional space, the couplings of $\Omega$ with the SM particles can become non-universal. Here we concentrate on the following 3-form field coupling to the SM muon leptons~\cite{Wu:2017urh,Wu:2018xah,Sezgin:1979zf}~\footnote{Note that, in Eq.~(\ref{LagHUFT}), the convention for the $\Omega$ coupling $g_h$ to muons differs from that given in Fig.~4 of Ref.~\cite{Wu:2018xah} by a factor $1/4$. }
\begin{eqnarray}\label{LagHUFT}
	{\cal L}_\Omega \supset -\frac{1}{2\cdot 4!} \Omega^{\mu\nu\rho\sigma} \Omega_{\mu\nu\rho\sigma}- \frac{m_\Omega^2}{2\cdot 3!} \Omega_{\mu\nu\rho} \Omega^{\mu\nu\rho} - i g_h \bar{\mu} \gamma^{\mu\nu\rho} \mu \Omega_{\mu\nu\rho}\,,
\end{eqnarray}     
where $\Omega_{\mu\nu\rho\sigma}$ denotes the four-form field strength of $\Omega_{\mu\nu\rho}$ defined by
\begin{eqnarray}
	\Omega_{\mu\nu\rho\sigma} \equiv \partial_{\mu} \Omega_{\nu\rho\sigma}-\partial_\nu \Omega_{\rho\sigma\mu} + \partial_\rho \Omega_{\sigma \mu\nu} - \partial_\sigma \Omega_{\mu\nu\rho}\,,
\end{eqnarray}
while $\gamma^{\mu\nu\rho}$ can be written as the combination of $\gamma$-matrices as follows
\begin{eqnarray}
	\gamma^{\mu\nu\rho} &\equiv& \gamma^{[\mu}\gamma^\nu\gamma^{\rho]} = \frac{1}{3!} \left(\gamma^\mu \gamma^\nu \gamma^\rho - \gamma^\nu \gamma^\mu \gamma^\rho + \gamma^\nu \gamma^\rho \gamma^\mu -\gamma^\rho\gamma^\nu \gamma^\mu + \gamma^\rho \gamma^\mu \gamma^\nu - \gamma^\mu \gamma^\rho \gamma^\nu\right) \nonumber\\
	&=& i\epsilon^{\mu\nu\rho\sigma} \gamma_\sigma \gamma^5\,,
\end{eqnarray}
where $\gamma^5 \equiv i\gamma^0 \gamma^1\gamma^2\gamma^3 = -i\epsilon^{\mu\nu\rho\sigma} \gamma_\mu \gamma_\nu \gamma_\rho \gamma_\sigma / (4!)$.

In Eq.~(\ref{LagHUFT}), we have included a mass term to the 3-form field $\Omega$, which explicitly breaks the original gauge symmetry $\Omega_{\mu\nu\rho} \to \Omega_{\mu\nu\rho} +3 \partial_{[\mu} b_{\nu\rho]}$ with $b$ a two-form gauge parameter field. Here we do not try to specify the possible origin of this mass term. Rather, we would like to emphasize its importance to the theory. Firstly, note that, without this mass term, the gauge symmetry involving $b$ is recovered in this Lagrangian of $\Omega$. Thus, the lightness of $\Omega$ is technically natural~\cite{tHooft:1979rat}, which can be understood in connection with this underlying broken gauge symmetry. Secondly, further significance of this mass term can be seen by analyzing its particle content in $\Omega$. In the massless case, $\Omega$ does not possess any physical propagating degree of freedom (dof) in four spacetime dimensions~\cite{Sezgin:1980tp}. To see this, let us divide the components of $\Omega$ into two classes, $\Omega_{0ij}$ and $\Omega_{ijk}$ with the Latin indices denoting the spatial dimensions. On the one hand, due to the antisymmetric nature of the field strength, the Lagrangian in Eq.~(\ref{LagHUFT}) cannot give rise to the kinetic term for $\Omega_{0ij}$ {\it {i.e.}}, $\partial_0 \Omega_{0ij}$ cannot exist. On the other hand, $\Omega_{ijk}\equiv \epsilon_{ijk} \phi$, actually being a pseudo-scalar in nature, suffers from the gauge transformation as $\Omega_{ijk} \to \Omega_{ijk} +3 \partial_{[i} b_{jk]}$, so that it is unphysical either. However, in the model with a massive 3-form field, the gauge invariance of $\Omega$ is explicitly broken, so that $\Omega_{ijk}$ retains as true physical propagating dof in the theory. Following the tensor-projection operator technique developed in Ref.~\cite{Rivers:1964,Sezgin:1979zf,Sezgin:1980tp,VanNieuwenhuizen:1973fi}, we can derive the propagator of the massive 3-form field $\Omega$ as follows
\begin{eqnarray}\label{PropOmega}
	i\Delta^\Omega_{\mu\nu\lambda,\,\rho\sigma\kappa} (k) = \frac{-i{\cal P}_{\mu\nu\lambda\,,\rho\sigma\kappa}(k)}{k^2-m_\Omega^2}\,,	
\end{eqnarray} 
where
\begin{eqnarray}\label{DefP}
	{\cal P}_{\mu\nu\lambda,\,\rho\sigma\kappa} (k) \equiv \frac{1}{6} \left[\theta_{\mu\rho} (\theta_{\nu\sigma}\theta_{\lambda\kappa} - \theta_{\nu\kappa}\theta_{\lambda\sigma}) - \theta_{\mu\sigma}(\theta_{\nu\rho}\theta_{\lambda\kappa}-\theta_{\nu\kappa}\theta_{\lambda\rho}) - \theta_{\mu\kappa}(\theta_{\nu\sigma}\theta_{\lambda\rho}-\theta_{\nu\rho}\theta_{\lambda\sigma}) \right]\,,
\end{eqnarray}
with
\begin{eqnarray}
	\theta_{\mu\nu}(k) \equiv \eta_{\mu\nu} - \frac{k_\mu k_\nu}{m_\Omega^2}\,.
\end{eqnarray}

{It has been well-known that there are many problems in constructing an interacting theories for higher-spin fields in the flat spacetime~(see {\it e.g.}, Ref.~\cite{Bekaert:2010hw} for a recent review and references therein), which is hampered by various no-go theorems~\cite{Weinberg:1964ew,Coleman:1967ad,Haag:1974qh, Grisaru:1976vm,Weinberg:1980kq, Porrati:2008rm, Benincasa:2007xk} in the literature. One may evade these no-go theorems by introducing higher-derivative couplings~\cite{Metsaev:2005ar,Boulanger:2006gr,Boulanger:2008tg} or placing the theory on the AdS spacetime~\cite{Fradkin:1986qy,Fradkin:1987ks,Vasiliev:2011knf}. Provided that the 3-form field $\Omega$ transforms non-trivially under the Lorentz symmetry group, one may wonder if our interacting theory might also be constrained by these no-go theorems. Here we would like to argue that our model of $\Omega$ is not plagued by these problems. First of all, as shown before, the massive 3-form field actually contains a single spin-0 pseudoscalar, rather than any higher-spin particles. Secondly, the previous no-go theorems focused on the massless higher-spin case, while $\Omega$ is dynamical only when it becomes massive. Thus, $\Omega$ may be decoupled when the energy is lower than its mass, and does not affect the infrared behavior of the theory. The above two reasons make either the Weinberg low-energy theorem~\cite{Weinberg:1964ew} or the Weinberg-Witten no-go theorem~\cite{Weinberg:1980kq} not be applicable to the present interacting 3-form model. It has been shown in the HUFT~\cite{Wu:2015wwa,Wu:2017rmd, Wu:2017urh, Wu:2018xah, Wu:2021ign,Wu:2021ucc,Wu:2022mzr} that the no-go theorems can also be avoided by introducing the concept of biframe spacetime. }

\section{The Muon $g-2$ Contribution}\label{SecG2}
\begin{figure}[!htb]
	\centering
	\hspace{-5mm}
	\includegraphics[width=0.5\linewidth]{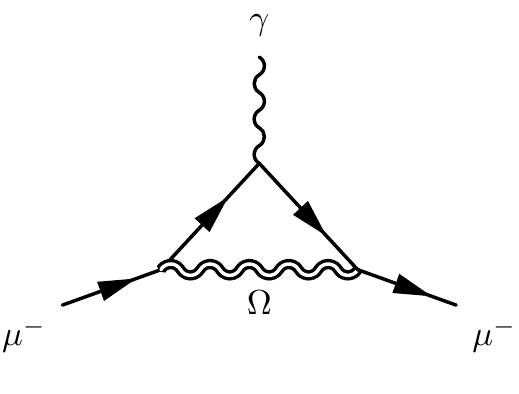}
	\caption{One-loop Feynman diagrams that would contribute to the muon $g-2$.}\label{Fig1Loop}
\end{figure}

Now it is ready to compute the novel contribution to the muon anomalous magnetic moment induced by $\Omega$. The relevant Feynman diagram is shown in Fig.~\ref{Fig1Loop}, where the double-wiggle line represents the massive 3-form field. 
By applying the Feynman rules according to the Lagrangian in Eq.~(\ref{LagHUFT}), the corresponding amplitude of Fig.~\ref{Fig1Loop} is given by
\begin{eqnarray}
	i{\cal M} &=& \int\frac{d^4 l}{(2\pi)^4} \left(g_h \gamma^{\rho\sigma \kappa}\right) \frac{i}{\slashed{l}+\slashed{k}_2 - m_\mu} \left(-ieQ_\mu \gamma^\mu\right) \frac{i}{\slashed{l}+\slashed{k}_1 -m_\mu} \left(g_\mu \gamma^{\tau\nu\lambda}\right) \frac{-i{\cal P}_{\tau\nu\lambda,\, \rho\sigma\kappa}(l)}{l^2 -m_\Omega^2} \nonumber\\
	&=& {g_h^2 eQ_\mu} \int\frac{d^4 l}{(2\pi)^4} \frac{\gamma^{\rho\sigma\kappa}[\slashed{l}+\slashed{k}_2 + m_\mu]\gamma^\mu [\slashed{l}+\slashed{k}_1 + m_\mu]\gamma^{\tau\nu\lambda} {\cal P}_{\tau\nu\lambda,\,\rho\sigma\kappa}(l)}{[(l+k_2)^2-m_\mu^2][(l+k_1)^2-m_\mu^2](l^2-m_\Omega^2)}\,,
\end{eqnarray}
$Q_\mu = -1$ and $m_\mu$ stand for the muon electric charge and mass, respectively.
By completing the square in the denominator with the Feynman parameters, the denominator can be transformed into  
\begin{eqnarray}
 &&	\frac{1}{[(l+k_2)^2-m_\mu^2][(l+k_1)^2-m_\mu^2](l^2-m_\Omega^2)} = \int^1_0 dx dy \frac{\Gamma(3)}{[l^2 + 2 l\cdot(x k_1 + y k_2)-(1-x-y)m_\Omega^2]^3} \nonumber\\
	&=& \Gamma(3) \int^1_0 dx dy \frac{1}{[(l+x k_1 + yk_2)^2 - \Delta]^3}\,,
\end{eqnarray}
where
\begin{eqnarray}
	\Delta \equiv (1-x-y)m_\Omega^2 + (x+y)^2 m_\mu^2 -xy q^2\,,
\end{eqnarray}
where $q\equiv k_2-k_1$ is the external photon momentum flowing into the vertex. By shifting the loop momentum as $l \to l-xk_1 -yk_2$, we can transform the above loop integral into the following form
\begin{eqnarray}
	i{\cal M} &=& eQ_\mu  g_h^2 \Gamma(3) \int dx dy \int \frac{d^4 l}{(2\pi)^4} {\cal P}_{\tau \nu\lambda,\, \rho\sigma \kappa} (l-xk_1 - yk_2) \nonumber\\
	&&\frac{\gamma^{\rho \sigma\kappa} [\slashed{l}-x\slashed{k}_1 + (1-y)\slashed{k}_2 + m_\mu]\gamma^\mu [\slashed{l}+(1-x)\slashed{k}_1 - y\slashed{k}_2 + m_\mu] \gamma^{\tau\nu\lambda}}{[l^2-\Delta]^3} \,.
\end{eqnarray}
We now contract the Lorentz indices and classify the obtained terms according to their divergence degrees. As a result, we find that the terms possibly proportional to the factor $m_\mu(-i\sigma^{\mu\nu}q_\nu)$ are logarithmically divergent at most, which indicates that those terms would give rise to the dominant contribution to the muon $(g-2)$. Thus, we firstly focus on these logarithmically divergent terms, which, after some tedious computations, gives
\begin{eqnarray}
	i{\cal M}^{\rm log} = e Q_\mu  g_h^2 m_\mu (-i\sigma^{\mu\nu}q_\nu) \int_0^1 dx dy \frac{6\Gamma(3)}{m_\Omega^2}  2\left[2(x+y)^2 - 6(x+y)+3 \right] {\cal I}_0(\Delta)\,,
\end{eqnarray}
where ${\cal I}_0(\Delta)$ is a logarithmically divergent integral regularized with the dimensional regularization~\cite{tHooft:1972tcz}:
\begin{eqnarray}
	{\cal I}_0 (\Delta) = \frac{i}{16\pi^2} \left[\frac{2}{\epsilon} - \gamma + \log\frac{\mu^2}{\Delta} \right]\,,
\end{eqnarray}
where $\epsilon \equiv 4-d$, and $\mu^2$ is a reference scale. 
It can be shown that the Feynman parameter integral over the UV divergent part of this amplitude vanishes 
\begin{eqnarray}
	\int^1_0 dx \int^{1-x}_0 dy [2(x+y)^2 -6(x+y)+3] =0\,.
\end{eqnarray}
Therefore, the one-loop contribution to the muon $g-2$ induced by $\Omega$ is finite. By further considering the large mass hierarchy between $\Omega$ and muons, we expand $\Delta$ in terms of $m_\mu^2/m_\Omega^2$ and just keep the leading-order term. Thus, the remaining part of $i{\cal M}^{\rm log}$ is of ${\cal O}(m_\mu q/m_\Omega^2)$, with the result given by
\begin{eqnarray}
	i{\cal M}^{\rm log} &=& eQ_\mu g_h^2 m_\mu (-i\sigma^{\mu\nu}q_\nu) \int^1_0 dx \int^{1-x}_0 dy \frac{12\Gamma(3)}{m_\Omega^2} [2(x+y)^2 - 6(x+y)+3] \times \nonumber\\
	 && (-1)\frac{i}{16\pi^2} \log(1-x-y) \nonumber\\
	 &=& i \frac{e}{2m_\mu} (i\sigma^{\mu\nu}q_\nu) \frac{18}{16\pi^2} \frac{Q_\mu g_h^2 m_\mu^2}{m_\Omega^2}\,.
\end{eqnarray}
The above calculation shows that the $\Omega$-induced anomalous contribution to the muon $g-2$ is dominated by terms of ${\cal O}(m_\mu^2/m_\Omega^2)$, so that we need to further consider the associated terms in the finite integrals. It can be shown by explicit calculations that these finite terms are given by
\begin{eqnarray}
	i{\cal M}^{\rm fin} &=& e Q_\mu g_h^2 m_\mu (-i\sigma^{\mu\nu}q_\nu)  \Gamma(3) \int^1_0 dx dy  \left(-\frac{i}{16\pi^2}\right) \frac{1}{2} (-12) \left[\frac{(x+y)^2-5(x+y)+4}{\Delta}\right]\nonumber\\
	&\simeq & i eQ_\mu g_h^2 m_\mu (i\sigma^{\mu\nu}q_\nu) \frac{\Gamma(3)(-12)}{2(16\pi^2)m_\Omega^2} \int^1_0 dx \int^{1-x}_0 dy (4-x-y) \nonumber\\
	&=&  i\frac{e}{2m_\mu} (i\sigma^{\mu\nu}q_\nu) \frac{40}{16\pi^2} \frac{(-Q_\mu)g_h^2 m_\mu^2}{m_\Omega^2}\,,
\end{eqnarray} 
where we also keep terms up to ${\cal O}(q m_\mu/m_\Omega^2)$ in the amplitude.

Note that the amplitude related to the observable muon $g-2$ is usually parameterized as follows~\cite{Peskin:1995ev}
\begin{eqnarray}\label{g2Def}
	i{\cal M} (\gamma \mu \to \mu)= \bar{u}(k_2) (-iQ_\mu) \left[e\gamma_\mu F_1(q^2) + \frac{ie \sigma_{\mu\nu} q^\nu}{2m_\mu} F_2(q^2)\right] u(k_1)\,,
\end{eqnarray}
where the measured muon anomalous magnetic moment can be identified as $\Delta a_\mu = F_2(0)$. In comparison with this definition, the leading-order result of the muon $g-2$ generated by $\Omega$ is given by
\begin{eqnarray}\label{g2Omega}
	\Delta a_\mu^{\Omega} = \Delta a^{\rm log}_\mu + \Delta a^{\rm fin}_{\mu} = \frac{11 g_h^2}{8\pi^2} \frac{m_\mu^2}{m_\Omega^2}\,.
\end{eqnarray}

\section{Constraints from Perturbativity and Unitarity}\label{SecUnitary}
The Lagrangian of the 3-form field $\Omega$ in Eq.~(\ref{LagHUFT}) suffers from the theoretical constraints such as perturbativity and unitarity. As a renormalizable dimension-4 operator, the effective interaction of $\Omega$ with muons should admit the usual perturbativity constraint. Note that the one-loop correction to this vertex can be estimated as $\delta g_h \sim g_h^3/(4\pi)^2$. The validity of perturbative expansion requires this one-loop correction be smaller than the tree-level coupling $g_h$, which can be transformed into the upper bound on $g_h$ as $|g_h| \lesssim 4\pi$~\cite{Nebot:2007bc}. 

\begin{figure}[!htb]
	\centering
	\hspace{-5mm}
	\includegraphics[width=0.49\linewidth]{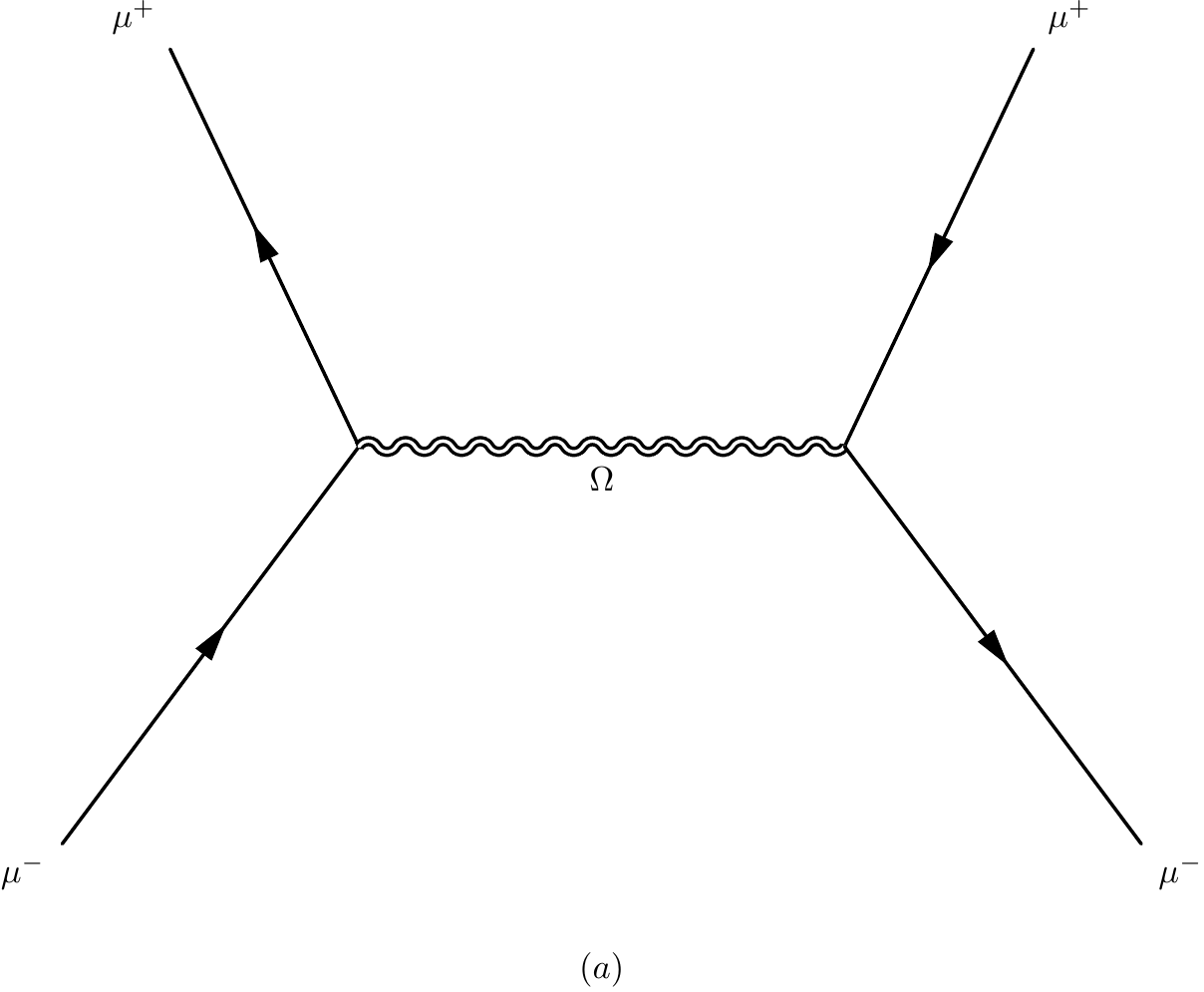}
	\includegraphics[width=0.49\linewidth]{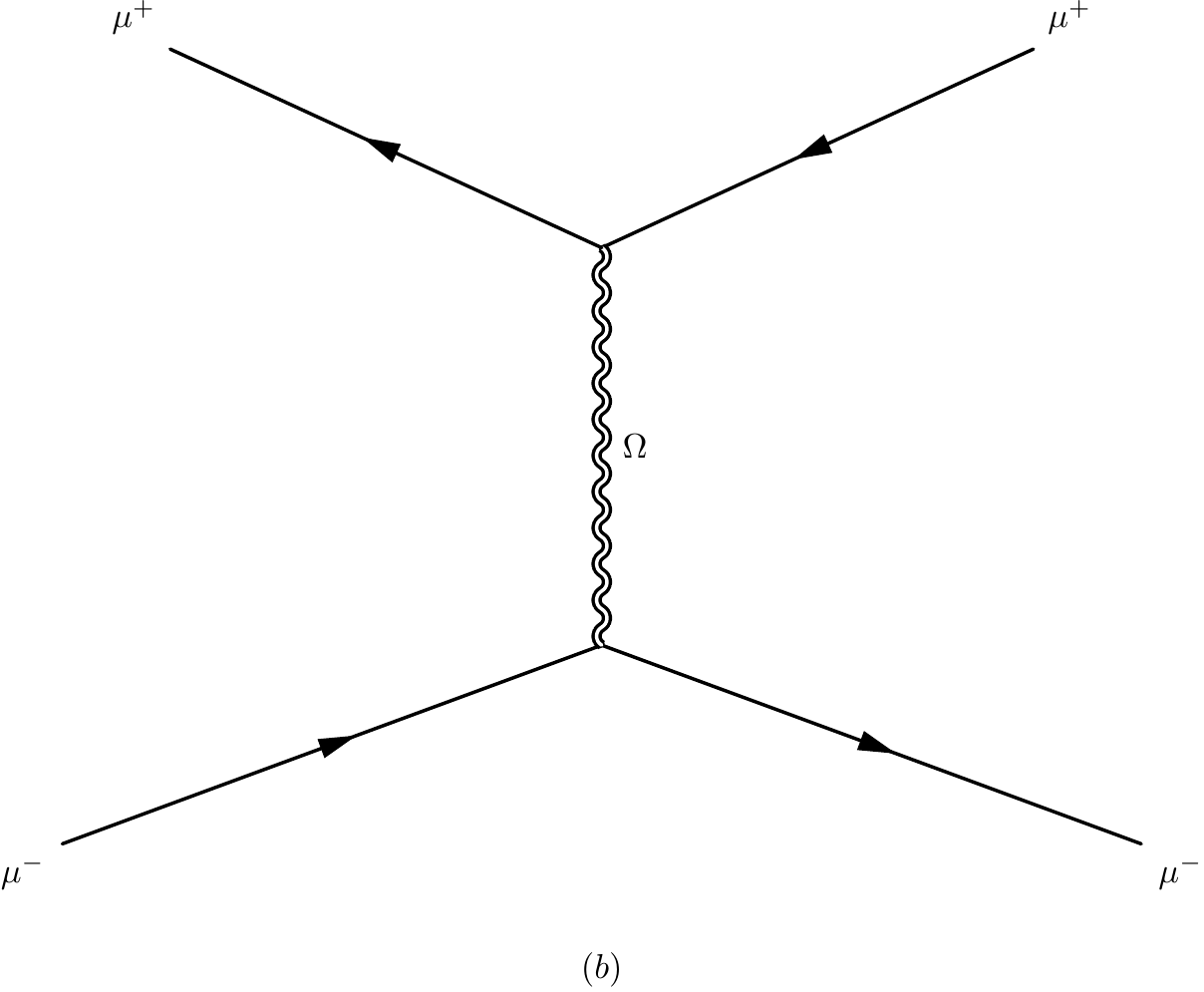}
	\caption{Feynman diagrams which give rise to the scattering process $\mu^- \mu^+ \to \mu^- \mu^+$: (a) $s$-channel; (b) $t$-channel.}\label{FigMM}
\end{figure}

Another restriction on the effective action of $\Omega$ is provided by the perturbative unitarity~\cite{Gell-Mann:1969cuq,Weinberg:1971fb,Lee:1977yc,Lee:1977eg} (also see Refs.~\cite{Glashow:1976nt, Huffel:1980sk, Maalampi:1991fb, Kanemura:1993hm, Akeroyd:2000wc, Das:2015mwa, Kanemura:2015ska, Goodsell:2018tti, Appelquist:1987cf, Chaichian:1987zt, Falkowski:2016glr,Banta:2021dek,SekharChivukula:2019yul,Chivukula:2020hvi,Chivukula:2021xod,Chivukula:2022tla,Huang:2022zop} for an incomplete list for the application of unitarity bounds in Beyond-SM theories). In order to obtain these unitarity bounds, we shall compute amplitudes of $\mu^- \mu^+$ elastic 2-to-2 scatterings shown in Fig.~\ref{FigMM}. 
By explicit calculations of these Feynman diagrams, the amplitudes of $s$- and $t$-channels are given by
\begin{eqnarray}\label{AmpST}
	i{\cal M}_s &=& \left[ \bar{v}(p_2) (g_h \gamma^{\rho\sigma\kappa}) u(p_1)\right] \frac{-i{\cal P}_{\rho\sigma\kappa,\,\mu\nu\lambda}(Q)}{Q^2-m_\Omega^2} \left[\bar{u}(k_1) (g_h \gamma^{\mu\nu\lambda}) v(k_2)\right]\nonumber\\
	&=& \frac{i 6 g_h^2}{m_\Omega^2} [\bar{u}(k_1) \gamma^\mu \gamma^5 v(k_2)] [\bar{v}(p_2) \gamma_\mu \gamma^5 u(p_1)]\,, \nonumber\\
	i{\cal M}_t &=&  \left[ \bar{u}(k_1) (g_h \gamma^{\rho\sigma\kappa}) u(p_1)\right] \frac{-i{\cal P}_{\rho\sigma\kappa,\,\mu\nu\lambda}(q)}{q^2-m_\Omega^2} \left[\bar{v}(p_2) (g_h \gamma^{\mu\nu\lambda}) v(k_2)\right]\nonumber\\
	&=& \frac{i 6 g_h^2}{m_\Omega^2} [\bar{u} (k_1) \gamma^\mu \gamma^5 u(p_1)] [\bar{v}(p_2) \gamma_\mu \gamma^5 u(k_2)]\,, 
\end{eqnarray}   
where we have defined $Q\equiv p_1 + p_2 = k_1 + k_2$ and $q\equiv p_1-k_1 = k_2-p_2$ and used the on-shell conditions for external muons. Also, we have worked in the high-energy limit $s\equiv Q^2 \gg m_\mu^2$ so that the muon mass can be ignored. In order to obtain the explicit expressions of amplitudes, we need to specify the helicty states for the four external muons. Before making explicit calculations, we should mention that the Lagrangian for $\Omega$ in Eq.~(\ref{LagHUFT}) is parity-invariant ($P$-invariant), so that two processes related by the parity transformation should have exactly the same amplitude. For example, we can have the following identities:
\begin{eqnarray}
	{\cal M} (\mu^-_R \mu^+_L \to \mu^-_R \mu^+_L) = {\cal M} (\mu^-_L \mu^+_R \to \mu^-_L \mu^+_R)\,,~
	{\cal M} (\mu^-_R \mu^+_L \to \mu^-_L \mu^+_R) = {\cal M} (\mu^-_L \mu^+_R \to \mu^-_R \mu^+_L)\,,
\end{eqnarray}
and so on. Therefore, the number of independent non-vanishing amplitudes are greatly reduced. In the following, we provide the calculation details for various nonzero independent amplitudes.

\subsection{$\mu^-_R \mu^+_L \to \mu^-_R \mu^+_L$} 
In the center-of-mass (com) frame, the momenta and spinors of the incoming particles are given by
\begin{eqnarray}\label{InCome}
  \mu^-_R: ~p_1 = (E, 0, 0, E)\,,&\quad& \mu^+_L: ~ p_2  = (E,0,0,-E)\,,\nonumber\\
   u_R(p_1) = \sqrt{2E} \left(\begin{array}{c}
  	0\\
  	0\\
  	1\\
  	0
  \end{array}\right)\,, &\quad& v_L(p_2) = \sqrt{2E} \left(\begin{array}{c}
  0\\
  0\\
  0\\
  -1
  \end{array}\right) \,,
\end{eqnarray}
while the outgoing particles are specified by
\begin{eqnarray}\label{OutGo}
	\mu_R^-: ~ k_1 = (E, E\sin\theta, 0, E\cos\theta)\,,&\quad& \mu_L^+:~ k_2 = (E,-E\sin\theta, 0, -E\cos\theta)\,, \nonumber\\
	  u_R(k_1) =  \sqrt{2E} \left(\begin{array}{c}
	  	0\\
	  	0\\
	  	\cos\frac{\theta}{2}\\
	  	\sin\frac{\theta}{2}
	  \end{array}\right)\,, &\quad& v_L(k_2) = \sqrt{2E} \left(\begin{array}{c}
	  	0\\
	  	0\\
	  	\sin\frac{\theta}{2}\\
	  	-\cos\frac{\theta}{2}
	  \end{array}\right) \,,
\end{eqnarray}
where we have taken the high-energy limit so that external muon masses can be neglected. By putting these explicit spinors into the general $\mu^-$$\mu^+$ elastic scattering amplitudes in Eq.~(\ref{AmpST}) for the $s$- and $t$-channels, we can obtain the total amplitude as follows
\begin{eqnarray}\label{ampRL}
	i{\cal M} (\mu^-_R \mu^+_L \to \mu^-_R \mu^+_L) = i{\cal M}_s + i {\cal M}_t = -24 i g_h^2 \left(\frac{s+t}{m_\Omega^2}\right) \,,
 \end{eqnarray}  
where we have used the definitions $s = Q^2$ and $t = q^2$.

\subsection{$\mu_R^- \mu_L^+ \to \mu^-_L \mu^+_R$}
In this case, the momenta for external muons are still taken as in Eqs.~(\ref{InCome}) and (\ref{OutGo}). Since the incoming $\mu^-\mu^+$ takes the same helicity configuration, the corresponding spinors are kept to be the same. However, due to the helicity flips for the outgoing muon pair, the spinors for these two particles are modified as follows 
\begin{eqnarray}
  u_L(k_1) =  \sqrt{2E} \left(\begin{array}{c}
	-\sin\frac{\theta}{2}\\
	\cos\frac{\theta}{2}\\
	0\\
	0
\end{array}\right)\,, &\quad& v_R(k_2) = \sqrt{2E} \left(\begin{array}{c}
	\cos\frac{\theta}{2}\\
	\sin\frac{\theta}{2}\\
	0\\
	0
\end{array}\right) \,.
\end{eqnarray}
It turns out that the $t$-channel amplitude in Eq.~(\ref{AmpST}) vanishes which is caused by the helicity mismatching in the spinor bilinears. On the other hand, the $s$-channel does contribute to the total amplitude as follows
\begin{eqnarray}\label{ampLR}
	i{\cal M} (\mu^-_R \mu^+_L \to \mu^-_L \mu^+_R) = i{\cal M}_s = 12ig_h^2 \left(\frac{t}{m_\Omega^2}\right) \,.
\end{eqnarray}

\subsection{$\mu^-_R \mu^+_R \to \mu^-_R \mu^+_R$}
In the com frame, the spinors for incoming and outgoing particles can be given by
\begin{eqnarray}
	u_R(p_1) = \sqrt{2E} \left(\begin{array}{c}
		0\\
		0\\
		1\\
		0
		\end{array}\right) \,,&\quad& v_R(p_2) = \sqrt{2E} \left(\begin{array}{c}
		1\\
		0\\
		0\\
		0
		\end{array}\right)\,,\nonumber\\
	u_R(k_1) = \sqrt{2E} \left(\begin{array}{c}
		0\\
		0\\
		\cos\frac{\theta}{2}\\
		\sin\frac{\theta}{2}
	\end{array}\right) \,,&\quad& v_R(k_2) = \sqrt{2E} \left(\begin{array}{c}
		\cos\frac{\theta}{2}\\
		\sin\frac{\theta}{2}\\
		0\\
		0
	\end{array}\right)\,.
\end{eqnarray} 
By exploiting the above explicit expressions of external muon spinors in Eq.~(\ref{AmpST}), we can yield the amplitudes for $s$- and $t$-channels. As a result, it is found that the $s$-channel ceases to contributing to the amplitude, while the $t$-channel gives the dominant contribution as follows
\begin{eqnarray}\label{ampRR}
	i{\cal M} (\mu^-_R \mu^+_R \to \mu^-_R \mu^+_R) = i{\cal M}_t = -12ig_h^2 \left(\frac{s}{m_\Omega^2}\right) \,.
\end{eqnarray}

\subsection{Numerical Analysis of Unitarity Bounds}
Due to the parity symmetry of the Lagrangian in Eq.~(\ref{LagHUFT}), Eqs.~(\ref{ampLR}), (\ref{ampRL}) and (\ref{ampRR}) give all the nonzero independent amplitudes for the process $\mu^-\mu^+ \to \mu^-\mu^+$, which can be used to yield the unitarity bounds on the 3-form field coupling to muons. Note that the unitarity of the S-matrix gives the following constraint~\cite{Lee:1977eg,Goodsell:2018tti,Banta:2021dek,Huang:2022zop} on the $s$-wave projected amplitude
\begin{eqnarray}\label{UniBound}
	\mbox{Re} (a_0) (\sqrt{s}) \leqslant \frac{1}{2}\,,
\end{eqnarray}
where $a_0(\sqrt{s})$ is defined as 
\begin{eqnarray}
	a_0(\sqrt{s}) = \sqrt{\frac{4|{\bf p}_i| |{\bf p}_f|}{2^{\delta_i+\delta_f}s}} \frac{1}{32\pi} \int^1_{-1} d(\cos\theta) {\cal M}(i \to f) = \sqrt{\frac{4|{\bf p}_i| |{\bf p}_f|}{2^{\delta_i+\delta_f}s}} \frac{1}{16\pi} \int^0_{-s} \frac{dt}{s} {\cal M}(i \to f)\,,
\end{eqnarray}
where the indices $\delta_{i(f)} = 1$ if the two particles in the initial (final) states are identical to one another, otherwise $\delta_{i(f)} = 0$. 

By making use of the definition of $a_0(\sqrt{s})$, we can obtain the $s$-wave projected amplitudes for different helicity configurations as follows
\begin{eqnarray}
	|a_0(\mu_R^- \mu_L^+ \to \mu_R^- \mu_L^+)| &=&  \frac{3 g_h^2}{4\pi} \frac{s}{m_\Omega^2} \sim \frac{3 g_h^2}{\pi} \frac{\Lambda^2}{m_\Omega^2}\,,\nonumber\\
	|a_0(\mu_R^- \mu_L^+ \to \mu_L^- \mu_R^+)| &=& \frac{3 g_h^2}{8\pi} \frac{s}{m_\Omega^2} \sim \frac{3 g_h^2}{2\pi} \frac{\Lambda^2}{m_\Omega^2} \,,\nonumber\\
	|a_0(\mu^-_R \mu^+_R \to \mu^-_R \mu^+_R)| &=& \frac{3 g_h^2}{4\pi} \frac{s}{m_\Omega^2} \sim \frac{3 g_h^2}{\pi} \frac{\Lambda^2}{m_\Omega^2}\,,
\end{eqnarray}
where in the last equality of each equation above we have taken the high-energy limit $s\sim 4\Lambda^2$ with $\Lambda$ as the UV cutoff scale. By further requiring the upper limit on ${\rm Re}(a_0)(\sqrt{s})$, we can obtain the following unitarity bound for each channel
\begin{eqnarray}
	\mu^-_R \mu^+_L \to \mu^-_R \mu^+_L / \mu^-_R \mu^+_R \to \mu^-_R \mu^+_R: &\quad& |g_h| \lesssim \sqrt{\frac{\pi}{6}} \frac{m_\Omega}{\Lambda} \,,\label{UniBoundEq}\\
	\mu^-_R \mu^+_L \to \mu^-_L \mu^+_R : &\quad& |g_h| \lesssim \sqrt{\frac{\pi}{3}} \frac{m_\Omega}{\Lambda}\,. 
\end{eqnarray}  
Therefore, it is found that the channels $\mu^-_R \mu^+_L \to \mu^-_R \mu^+_L$ and $\mu^-_R \mu^+_R \to \mu^-_R \mu^+_R$ provide the strongest constraint with $|g_h| \lesssim \sqrt{\pi/6} (m_\Omega/\Lambda)$. 

\section{Numerical Results}\label{SecResult}
\begin{figure}[!htb]
	\centering
	\hspace{-5mm}
	\includegraphics[width=0.48\linewidth]{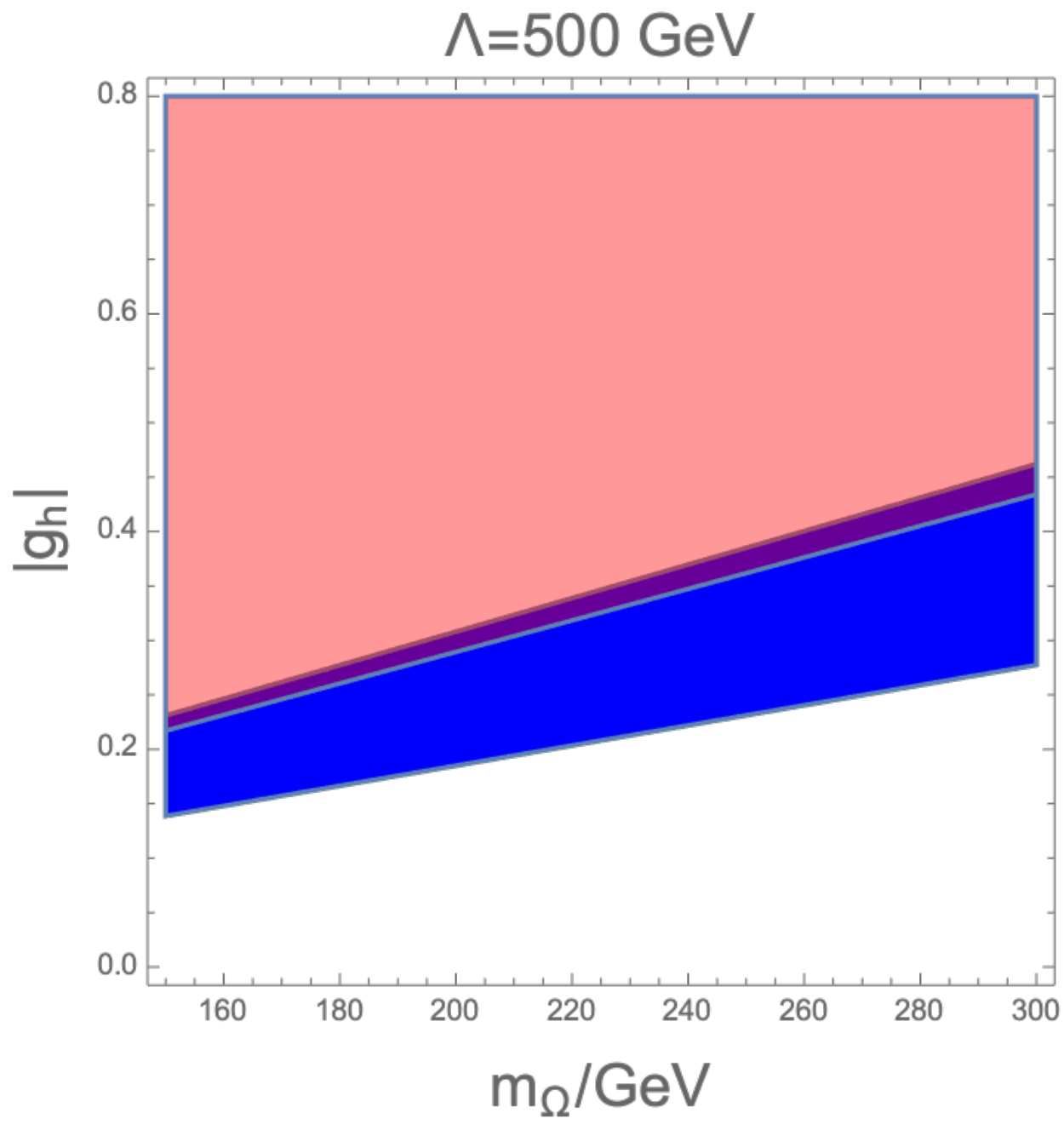}
	\includegraphics[width=0.48\linewidth]{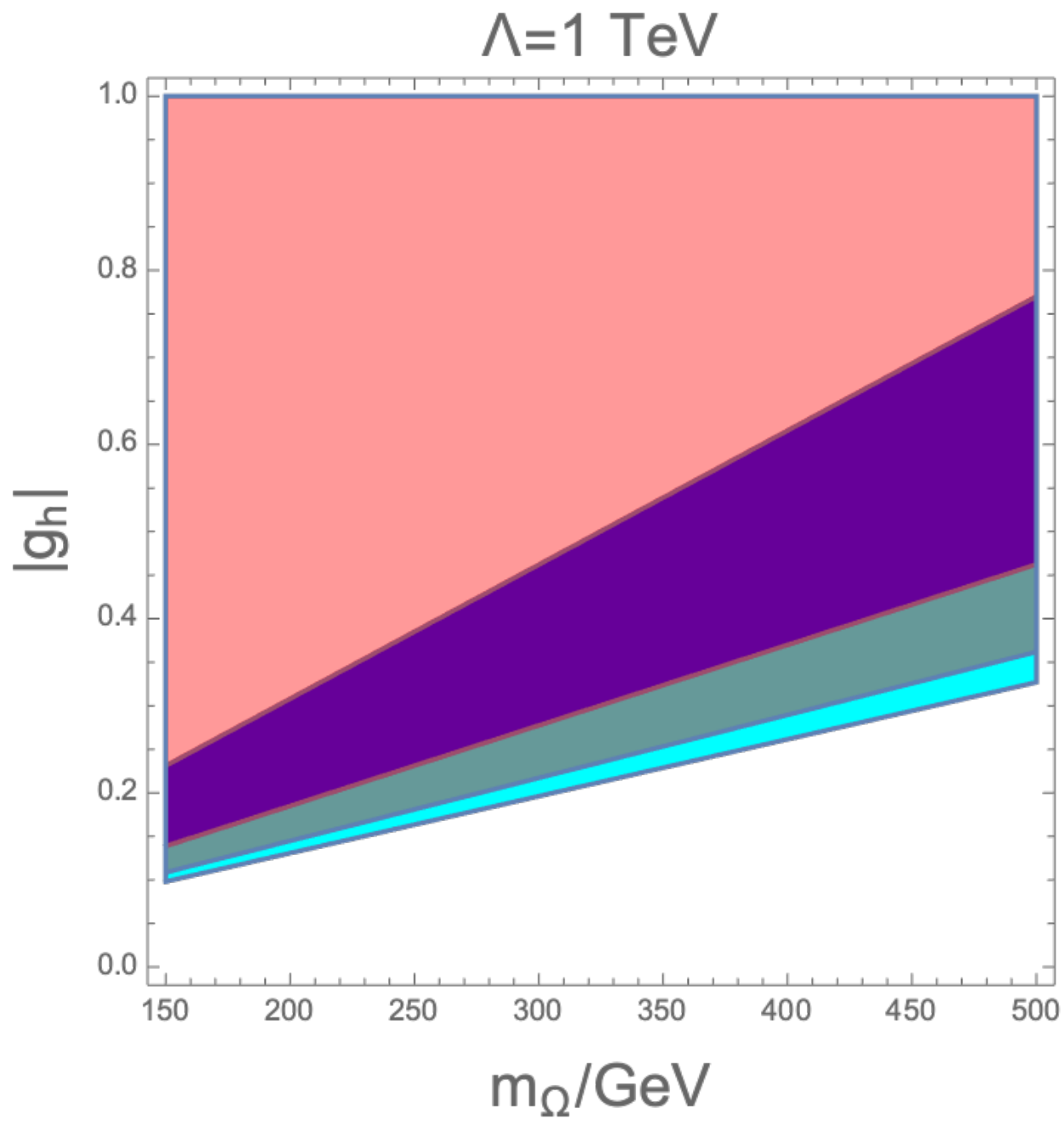}
	\caption{Parameter spaces on the $m_\Omega$-$g_h$ plane for the UV cutoff to be $\Lambda = 500$~GeV (left panel) and 1~TeV (right panel), respectively. In each plot, the blue region represents the parameter space which can explain the muon $g-2$ at $2\sigma$ CL, while the red region is excluded by the perturbative unitarity bound. The cyan region on the right panel can account for at least $50\%$ of the muon $g-2$ anomaly. }\label{plotRes}
\end{figure}

Given the muon $g-2$ expression in Eq.~(\ref{g2Omega}) and the strongest unitarity bound in Eq.~(\ref{UniBoundEq}), we are now ready to explore the parameter space of this model. The final numerical result is plotted on the $m_G$-$|g_h|$ plane in Fig.~\ref{plotRes}, where the cutoff scale entering the unitarity bound is chosen to be $\Lambda = 500$~GeV (left panel) and 1~TeV (right panel). The solid blue band in each plot corresponds to the parameter space explaining the muon $g-2$ data at $2\sigma$ CL, while the red shaded region is disfavored by unitarity. Here we do not show the theoretical limit from perturbativity $|g_h| \lesssim 4\pi$, since it is too weak to be useful in constraining this model. Moreover, in order to be consistent with the effective field theory treatment, it is usually required that the mass of $\Omega$ should be much smaller than the cutoff scale. Thus, we take the upper boundary of the 3-form field $\Omega$ mass to be $m_\Omega= 300$~GeV (500~GeV) for $\Lambda = 500$~GeV (1~TeV) in Fig.~\ref{plotRes}. 

In the left panel of Fig.~\ref{plotRes} with the cutoff scale $\Lambda = 500$~GeV, even though the unitarity bound begins to restrict the large coupling region, there is still a considerable amount of parameter spaces to fully explain the muon $g-2$ anomaly with 150~GeV~$\leqslant m_G \leqslant 300$~GeV and $0.1 \lesssim  |g_h|\lesssim 0.5$. On the other hand, when the cutoff scale is increased to $\Lambda \gtrsim 800$~GeV, it is found that the blue colored region totally resolving the muon $g-2$ discrepancy is ruled out by unitarity arguments, which is illustrated in the right panel of Fig.~\ref{plotRes} for $\Lambda = 1$~TeV. This implies that the current muon $g-2$ data favors a light tensor field $\Omega$ with $m_G \sim {\cal O}$(100~GeV) and a low cutoff scale $\Lambda \lesssim 800$~GeV. However, if we relax our requirement and only need our 3-form model to account for at least $50\%$ of the muon $g-2$ anomaly, more parameter space now opens, which is colored in cyan in the plot for $\Lambda = 1$~TeV. It is seen that some of the cyan region now escapes the strong unitarity bound, indicating that our model of the 3-form field can explain at least $50\%$ of the muon $g-2$ discrepancy.  


\section{Conclusions}\label{SecConclusion}
In the present work, we have considered the possibility to explain the long-standing muon $g-2$ anomaly by introducing in the SM a 3-form field $\Omega$, which can naturally appear in many fundamental theories such as the type IIA string theory and the HUFT. In particular, we have presented the leading-order analytic expression of the muon $g-2$ by calculating the one-loop Feynman induced by $\Omega$. We have also taken into account the theoretical constraints from perturbativity and unitarity. Especially, we have computed the independent tree-level $\mu^+\mu^-$ scattering amplitudes of all initial and final helicty configurations, from which we have seen that the channels $\mu^-_R \mu^+_L \to \mu^-_R \mu^+_L$/$\mu^-_R \mu^+_R \to \mu^-_R \mu^+_R$ lead to the most stringent bound on this 3-form field model. Then we have numerically explored the parameter space which is of interest to explain the muon $g-2$ anomaly.  As a result, despite the strong constraint from the unitarity, we have still found a substantial parameter region which can solve the muon $g-2$ discrepancy with the benchmark scenario as $m_\Omega \sim 250$~GeV, $|g_h| \sim 0.3$, and $\Lambda \sim 500$~GeV. However, when the UV cutoff scale has risen to a larger value $\Lambda \gtrsim 800$~GeV, all parameter regions fully resolving the $(g-2)_\mu$ anomaly have been excluded by unitarity. Nevertheless, in the case with $\Lambda = 1$~TeV, we have also shown that at least 50$\%$ of the muon $g-2$ discrepancy can be accounted for by our 3-form field theory without conflicting with the strong unitarity bound. 

Note that the interpretation of the muon $g-2$ anomaly requires the 3-form field $\Omega$ to be relatively light and to have a moderately strong coupling to $\mu$ leptons, so that it should be well tested by the existing collider experiments. For example, if we extend this model by assuming that $\Omega$ has a universal coupling to all the SM leptons and quarks, the effective $\Omega$-gluon interaction could be inevitably generated via SM quark loops. Thus, the 3-form particle $\Omega$ can be produced at hadronic colliders such as the LHC mainly through the gluon fusion channel. By observing its decay products such as $t\bar{t}$, dijet, diphoton, diboson, and dileptons, such the universal coupling model with a light $\Omega$ has already been ruled out by the current ATLAS~\cite{ATLAS:2018rvc,ATLAS:2018qto,ATLAS:2019nat,ATLAS:2019erb,Wang:2021rvc} and CMS~\cite{CMS:2018rkg, CMS:2018mgb, CMS:2019qem,CMS:2018ipm,CMS:2018dqv,Radburn-Smith:2018wfo} data. On the other hand, if this 3-form field is leptophilic, {\it i.e.}, $\Omega$ only couples to the SM lepton sector strongly, then the LHC constraints cannot be applied here. Nevertheless, with the assumption that $\Omega$ interacts with $e^\pm$ by a similar coupling to muons, such a model can be falsified in the near-future $e^+e^-$ colliders such as CLIC~\cite{Linssen:2012hp,Aicheler:2012bya}, ILC~\cite{Baer:2013cma}, CEPC~\cite{CEPCStudyGroup:2018rmc,CEPCStudyGroup:2018ghi} and FCC-ee~\cite{FCC:2018evy}. Note that the cross section of $e^- e^+ \to \mu^- \mu^+$ mediated by $\Omega$ is given by
\begin{eqnarray}\label{SigmaEE}
	\sigma (e^- e^+ \to \mu^- \mu^+) = \frac{3 g_h^4}{\pi} \frac{s}{m_\Omega^4}\,,
\end{eqnarray}
where $s$ is the com energy squared of the $e^+e^-$ pair. Interestingly, the cross section in Eq.~(\ref{SigmaEE}) shows a monotone increase with $s$, and the absence of any pole structure at $s = m_\Omega^2$, which is quite unusual in an $s$-channel process. Such a feature is caused by the fact that $(s-m_\Omega^2)$ originally appearing in the denominator of the $\Omega$ propagator in Eq.~(\ref{PropOmega}) has been canceled exactly by the same factor generated in the numerator of the $s$-channel amplitude. Furthermore, {if we take benchmark parameters as $m_\Omega = 250$~GeV, $g_h = 0.3$, and $\sqrt{s} \sim 300~{\rm GeV}$ which is well below the cutoff scale $\Lambda = 500$~GeV as indicated by the left panel of Fig.~\ref{plotRes}, the cross section for the $\mu^-\mu^+$ pair production induced by $\Omega$ can be estimated to be $\sim 0.7$~ab. }
Such a large cross section with its unconventional rise with $s$ indicates that the process $e^- e^+ \to \mu^- \mu^+$ is a promising target in searching for the 3-form field $\Omega$. 

{Finally, we would like to emphasize that our treatment of the 3-form field $\Omega$ is, at best, effective only at low energies. Indeed, as clearly seen from Eqs.~(\ref{ampRL}), (\ref{ampLR}), (\ref{ampRR}), and (\ref{SigmaEE}),  the $\mu^- \mu^+ \to \mu^- \mu^+$ amplitudes of various polarization configurations and the cross section of $e^- e^+ \to \mu^-\mu^+$ increases with the com energy squared $s$, and would break the unitarity at high energies. Such a pathology is why we have introduced a UV cutoff scale $\Lambda$ in the theory. It is highly possible that the theory is unitarized at energies around $\Lambda$ by some mechanisms, such as the Higgs or St$\ddot{\rm u}$ckelberg mechanisms, which might also give the mass to the 3-form field and make $\Omega$ a propagating degree of freedom. Or $\Omega$ is just an effective description of the dynamics when $s$ is much lower than $\Lambda$, as the massless counterpart is not dynamical at all. Another theoretical problem related to $\Omega$ is the renormalizability of the theory. As shown in Eqs.~(\ref{PropOmega}) and (\ref{DefP}), the propagator of the 3-form field $\Omega$ does not obey the conventional power counting rule as the SM scalar or vector particles, since adding more internal $\Omega$ propagators would even increase the UV divergences for the loop integrals, rather than ameliorating the UV behavior. However, in the present work, we concentrate on explaining the muon $g-2$ in terms of the one-loop effects of the 3-form field $\Omega$, whose finiteness reveals that its prediction is only determined by low-energy dynamics, regardless of the high-energy details or the UV completion of the theory.   }
 
 
\section*{Acknowledgements}
\noindent DH is supported in part by the National Natural Science Foundation of China (NSFC) under Grant No. 12005254, the National Key Research and Development Program of China under Grant No. 2021YFC2203003, and the Key Research Program of Chinese Academy of Sciences under grant No. XDPB15. YT is supported in part by NSFC under Grants No.~11851302, the Fundamental Research Funds for the Central Universities and Key Research Program of the Chinese Academy of Sciences. YLW is supported in part by the National Key Research and Development Program of China under Grant No.~2020YFC2201501, and NSFC under Grants No.~11851303, No.~11690022, No.~11747601, the Strategic Priority Research Program of the Chinese Academy of Sciences under Grant No. XDB23030100, and NSFC special fund for theoretical physics under Grant No. 12147103.

\appendix


\end{document}